*CCC Whitepaper*
December 2025

# The Imperative for Grand Challenges in Computing

### Authors

William Regli (University of Maryland), Rajmohan Rajaraman (Northeastern University), Daniel Lopresti (Lehigh University), David Jensen (University of Massachusetts, Amherst), Mary Lou Maher (Computing Research Association), Manish Parashar (University of Utah), Mona Singh (Princeton University), Holly Yanco (University of Massachusetts, Amherst)

## Introduction

Computing is now an indispensable component of nearly all technologies and is ubiquitous for vast segments of society. It is also essential to discoveries and innovations in most disciplines. However, while past grand challenges in science have *involved* computing as one of the tools to address the challenge, these challenges have not been principally *about* computing. Why has the computing community not yet produced challenges at the scale of grandeur that we see in disciplines such as physics, astronomy, or engineering? How might we go about identifying similarly grand challenges? What are the grand challenges of computing that transcend our discipline's traditional boundaries and have the potential to dramatically improve our understanding of the world and positively shape the future of our society?

We are seeking challenge problems that are sufficiently compelling as to both ignite the imagination of computer scientists and draw researchers from other disciplines to computational challenges. Climate, health, energy, national security, and the acceleration of scientific inquiry and economic well-being are some areas that provide a context for challenge problems in computing, but these domains are not, on their own, computing. It is our belief that multi-disciplinary problems necessitate the creation of use-inspired and translational research in computing that drives innovation in both computing and the problem domain as we search for solutions.

There is a significant benefit in us, as a field, taking a more intentional approach to "grand challenges." While the pace of development in computing over the past few decades has been exhilarating, too often the implications of our progress do not become apparent until after large investments have been committed and the impacts of these investments on science and the public may not be clear or as intended. We believe that taking a long view is appropriate as our field continues to mature.

This paper emphasizes the importance, now more than ever, of defining and pursuing grand challenges in computing as a field, and being intentional about translation and realizing its impacts on science and society. Building on lessons from prior grand challenges, the paper explores the nature of a grand challenge today emphasizing both scale and impact, and how the community may tackle such a grand challenge, given a rapidly changing innovation ecosystem in computing. The paper concludes with a call to action for our community to come together to define grand challenges in computing for the next decade and beyond.

## Why Grand Challenges for Computing and Why Now?

The invention of calculus and the establishment of the scientific method in the 17th century created the framework for the industrial revolution and the past three centuries of progress enabled by the field of engineering. This century is the century of computing machines. Computer science, which is now 70 years old, has firmly established a new framework that can enable us to reason about and address the challenges and opportunities facing society. Computing is not only an essential tool for science and engineering, but within the last 30 years — from the rise of the Internet, to access via smart phones, communications via social media, and now the widespread use of AI systems — computing is literally reprogramming our civilization and how it functions. We are in the midst of a massive transition in human society to one that runs on a computational substrate and we contend that computer science has not been asking enough big questions. What are the grand challenges uniquely centered on computing research that can catalyze the next generation of discoveries?

## A Brief History of Grand Challenges

The concept of "grand challenges" is not new and several have changed the course of history. In 1714, the British government established the Longitude Act,[1] offering a prize for a means of accurately calculating the longitude of ships at sea. The War of 1812 made the U.S. Army aware of the benefits of interchangeable parts, leading to several decades of innovation in measurement and manufacturing. In 1900, David Hilbert's stated 27 fundamental mathematical problems,[2] the topics of which became the focus of mathematics for over half of the 20th century. New York City hotelier Raymond Orteig[3] established a prize in 1919 for the first powered flight over the Atlantic Ocean — eventually won in 1927 by Charles Lindbergh. World

---

[1] Wikipedia, "Longitude Act," https://en.wikipedia.org/wiki/Longitude_Act.
[2] Wikipedia, "Hilbert's Problems," https://en.wikipedia.org/wiki/Hilbert%27s_problems.
[3] Wikipedia, "Orteig Prize," https://en.wikipedia.org/wiki/Orteig_Prize.

War II precipitated several technological challenges that led to innovations such as the proximity fuse,[4] radar,[5] and the atomic bomb. The Cold War gave rise to the space race and the Apollo Program. The concepts of "Manhattan Project" and "Moonshot" are now synonymous with our notion of a "Grand Challenge".

Recently, grand challenges in other areas of science and engineering have produced space telescopes,[6] the Human Genome Project,[7] and other large-scale scientific instruments.[8] The National Academy of Engineering organized a multi-year planning effort that culminated in 2013 with "14 Grand Challenges for Engineering for the 21st Century,"[9] which included topics such as "Manage the Nitrogen Cycle" and "Prevent Nuclear Terror." The federal government has often taken a leadership role in this process; recent examples include the 21st Century Grand Challenges,[10] the Cancer Moonshot in 2016, and the National Quantum Initiative Act of 2018.

## Past Computing Challenges

Computing, in various forms, is an essential element of many of these "grand challenges" (especially the recent ones). However, while computing technologies are essential to many of these challenges, nearly all are not principally about computing itself. What can we learn from past grand challenges efforts, specifically within the field of computing?

Initiated by Bob Kahn and others in the 1980s, DARPA's Strategic Computing Initiative (SCI)[11] outlined the framework for the computer science innovations and advancements that emerged in the subsequent decades. Initial "challenges" were related to autonomous driving, language understanding, and human-machine teaming (i.e., "the pilot's assistant"). These came to fruition in a variety of ways. DARPA's Grand Challenges in 2004, 2007, and 2012-2015 pushed the frontiers on fully autonomous ground vehicles and autonomous emergency-maintenance robots — technologies we now see in commercial practice. Apple's Siri[12] emerged from DARPA programs (most specifically the Cognitive Assistant that Learns and Organizes (CALO) program) in the early 2000s as did innovations in human language. The leadership mantles for these areas were assumed by industry, leading to numerous commercial innovations that are now ubiquitous.

---

[4] Wikipedia, "Proximity Fuze," https://en.wikipedia.org/wiki/Proximity_fuze.
[5] Wikipedia, "History of Radar," https://en.wikipedia.org/wiki/History_of_radar.
[6] Examples include major space observatories such as the Hubble Space Telescope (1970s–present), Kepler (2000–2018), Spitzer (1990–2020), and the James Webb Space Telescope (1996–present).
[7] The initiative — planned in 1984, launched in 1990, and completed in January 2022 — produced the first complete map of human DNA.
[8] Examples include the Large Hadron Collider (1998–2010) and the Laser Interferometer Gravitational-Wave Observatory (LIGO).
[9] Engineering Challenges, "Challenges," https://www.engineeringchallenges.org/challenges.aspx.
[10] Office of Science and Technology Policy (U.S.), "21st Century Grand Challenges," https://obamawhitehouse.archives.gov/administration/eop/ostp/grand-challenges.
[11] Wikipedia, "Strategic Computing Initiative," https://en.wikipedia.org/wiki/Strategic_Computing_Initiative; and Roland, Alex, and Philip Shiman, Strategic Computing: DARPA and the Quest for Machine Intelligence, 1983–1993, MIT Press, https://mitpress.mit.edu/9780262529266/strategic-computing/.
[12] Wikipedia, "Siri," https://en.wikipedia.org/wiki/Siri.

In 2002, the Computing Research Association (CRA) hosted a significant "grand challenges" conference[13] that was attended by many computing luminaries including several founders of the field of computer science. Contemporaneously, there were similar efforts in the United Kingdom to identify computing "grand challenges." CRA's 2002 conference produced five candidate challenges: (1) Systems you can count on; (2) A teacher for every learner; (3) 911.net (ubiquitous information systems); (4) Augmented cognition; and (5) Conquering complexity. Assessing progress since that time is difficult because, for the most part, these challenges lack specificity and metrics. In at least two of the cases, the objectives are entirely subjective, making it difficult to know exactly when one has been achieved.

In 2003, a second Grand Challenge Symposium was organized by CRA, this one focused on trustworthy computing. The conference brought together leaders in computer systems and security to identify ambitious challenges that define a success criteria, excite the imagination, and have potential for broad social impact. Four grand challenges were identified; in brief: (1) eradicate denial-of-service (DOS) attacks within a decade; (2) create principles, tools, and methods to operate critical infrastructure and support democratic institutions; (3) create a framework for end users to manage comprehensive security and privacy; and (4) in ten years, create and implement a framework to manage risk in IT systems comparable to quantitative financial risk-management.

A recent remarkable 20-year retrospective[14] of the 2003 CRA conference invited the attendees of the original event to reflect on the current status of the grand challenges proposed at the time. Their insightful report argues that these challenges have not been met, pointing to a range of possible reasons, including underestimation of the pace and scale of technological change and capabilities of threat actors, subjective formulation of some challenges, lack of understanding of application areas, and the sheer complexity of underlying problems. Nevertheless, the retrospective also highlights the considerable progress made in the pursuit of these grand challenges: for instance, cybersecurity becoming an integral part of systems and an "enabler" for designers, and the design of trustworthy systems from unsecure components.

---

[13] Computing Research Association, "CRA Grand Research Challenges 2002," https://archive.cra.org/Activities/grand.challenges/index.html.

[14] DeMillo, Richard A., and Eugene H. Spafford, "Grand Challenges in Trustworthy Computing at 20: A Retrospective Look at the Second CRA Grand Challenges Conference," Communications of the ACM, August 27, 2025, https://cacm.acm.org/research/grand-challenges-in-trustworthy-computing-at-20-a-retrospective-look-at-the-second-cra-grand-challenges-conference/.

## What Might We Observe About Previous Computing Challenges?

While change may seem to appear suddenly, computing revolutions often emerge over the course of several generations of foundational, use-inspired, and translational research activity, perhaps unfolding over many years until critical tipping points are reached. Representative examples include the Internet, human-computer interfaces, deep learning, large language models, various cybersecurity solutions, and open source software. This "gradual progress followed by explosion" seems to be a fundamental property of our discipline. The result is that we proceed along, building tools and systems, until some threshold of capabilities are reached that enables us to interconnect these tools and systems in a way that creates an unprecedented impact that was not necessarily envisioned at the outset.

While there may be some key actors, the conditions for success in computing have usually been the product of many individuals and organizations contributing multiple, varying disciplinary perspectives. The development of the ARPAnet and Internet through the 1970s, and the emergence of the World Wide Web in the late 1990s are illustrations of such collective innovations. Compelling applied and use-inspired objectives create the need to bring these emerging pieces together. Examples range from the development of IBM 360 for general purpose business computing to the "Mother of All Demos" of Douglas Engebart in 1968 showing integrated collaboration and software development, and the introduction of the iPhone in 2007 which highlights use-inspired objectives. Often, computing revolutions come from major *intentional* and *translational* investments in infrastructure, testbeds, and challenge problems to motivate and focus the community. Examples include DARPA's Strategic Computing Initiative of the 1980s and National Science Foundation's "10 Big Ideas," which often included shared resources and testbeds. The supercomputer and cyberinfrastructure facilities managed by the Department of Energy and the NSF serve a similar purpose.

In addition to SCI and CRA's efforts, there have been additional computational challenges sponsored by NSF, DARPA, and others, including those in focused topics such as cybersecurity, high-performance computing, and, of course, AI. What lessons might we learn from past efforts at "grand challenges?" We observe three things.

First, some begin with a profound scientific question for which the challenge is to design and build *a fundamentally new kind of scientific instrument* to understand and ultimately answer the question. This is seen most obviously in the various programs that have built telescopes and other observational capabilities (e.g., LIGO, JWSC, etc.). In computing, the goal of creating an operational Quantum Computer might be seen to be a similar example.

Second, most of these challenges (certainly including those that build scientific instruments) have *tangible and definite outcomes*. Like the Manhattan Project or the original Moonshot, challenges related to mathematical proof (Hilbert's questions and the Millennium Prize) certainly fall into this category. The same can also be said for the Human Genome Project, Operation Warp Speed for the development of the COVID-19 vaccines, and any other challenge with a specific engineering objective.

Lastly, an effective challenge, even one that lacks a tangible and definite goal, at least has *metrics* by which one can measure progress. Self-driving vehicles have seen a steady increase in the complexity of measurement: from driving across a desert, to passing the California driver's test, to fully autonomous driving. The current quest for artificial general intelligence (AGI) might also be viewed in this way, measured as successive improvements against various benchmarks—indeed, the very disagreement around measurements and metrics for AI systems may be a fundamental part of the challenge itself.

## Thinking About Grand Challenges in Computing

As we explore grand challenges in computing, it is essential to develop a framework for conceptualizing and advancing such challenges. Such a framework must enable holistic thinking about computing research from different perspectives; in addition to *scientific and societal* grand challenges where computing research is essential and a driving force, the framework should enable the formulation of grand challenges for *computer science as a field*, including the acceleration and amplification of its translational impacts. Toward this end, we posit that grand challenges must satisfy the following key criteria:

- **Impact:** The challenge should address a critical need or urgent societal problem.

- **Ambition:** The challenge should be ambitious and have the potential to drive significant progress in the computing field.

- **Feasibility:** The challenge should be beyond the reach of current computing techniques but perhaps barely feasible using disruptive new ideas on the horizon.

- **Interdisciplinarity:** The challenge should draw expertise from multiple computing research areas and stakeholders from diverse backgrounds.

- **Measurability:** Progress towards addressing the challenge should be measurable. While a formal quantification of progress may not always be possible at the stage of formulating the challenge, it is critical to articulate outcomes that will result from a successful completion of the challenge.

It is important that the process for formulating grand challenges brings together a broad spectrum of researchers, practitioners, and stakeholders, from within computing and beyond. Such an approach will create space for brainstorming on the challenges within the framework of the above criteria, and can identify grand challenge prototypes which may be expressed by brief answers to the following questions:[15]

- *What is the challenge?*

- *Why now and what makes this challenge "grand" and (barely) feasible?*

- *How will we know if we succeeded?*

## Where Are Today's Computing Grand Challenges?

Identifying grand challenges is a presumptuous task that requires an amount of hubris. The tendency is to gravitate toward topics and themes that we are familiar with — akin to those that might appear during the course of "normal computer science." Thomas Kuhn[16] famously described that scientific revolutions occur when "normal science" is breaking down, where instruments seem to fail and measurements are lacking, and when there are crises that must be solved. He notes that, while the paradigm shift created by revolutions may only be fully visible in hindsight, the broad outlines of the subject of revolution are often evident from the events of the day.

The Grand Challenges Task Force of the CCC Council attempted to get a pulse on where the crises and opportunities might lie today. We organized four virtual roundtable sessions in Spring 2024 with 36 attendees, which included researchers across the computer science spectrum. In July 2024, the Task Force led a dynamic 90-minute session at the CRA Summit in Snowbird with over 70 attendees, drawn from leadership in academia and industry research. More recently, in May 2025, the Task Force held a grand challenges session at the CIFellows Symposium with 40 early-career computing researchers. In parallel, the CCC's AI Futures Task Force also conducted a series of visioning exercises, including those at AAA2025. While a truly compelling grand challenge cannot be meaningfully identified in a short ideation session, perhaps we can infer the major vectors that form their principal components:

- "Fundamentally, computer science is a science of abstraction — creating the right model for thinking about a problem and devising the appropriate mechanizable techniques to solve it."[17] Various domains, from healthcare to physics, offer an endless supply of challenges that require computational abstraction.

---

[15] These three questions draw from the Heilmeier Catechism (https://www.darpa.mil/about/heilmeier-catechism) and from Arati Prabhakar's article "In the Realm of the Barely Feasible," *Issues in Science and Technology*, vol. 37, no. 1, 2020, pp. 34-40.
[16] Wikipedia, "The Structure of Scientific Revolutions," https://en.wikipedia.org/wiki/The_Structure_of_Scientific_Revolutions.

- We are changing the physics of how to compute and what our notion of "compute" actually is. The discipline is moving beyond binary arithmetic executed on transistors and switches to computing with qubits, photons, neurons and chemicals. Challenges in this category build on the notion that nature–and life itself–may be a large computing machine.

- Understanding the world is a common theme for possible challenges. Increasing the fidelity of modeling and simulation in our supercomputer centers, for example, improves our tools for simulating the physical world. Models can relate to physics, chemistry, health, climate, or social systems. Improving fidelity and scale by factors of 10x or 100x offer a variety of challenges—some of which are possibly grand.

- Physically sensing and interacting with the environment. The classic framework of an "intelligent agent" as a machine that perceives through sensors and then acts via effectors implies that improvement in the sense/act loop may offer challenges across a variety of domains including robotics.

- The role and potential of Artificial Intelligence is a ubiquitous theme. Herb Simon's "Sciences of the Artificial"[18] argues that "devising courses of action" is a primary activity that separates humans from the natural world. The natural world does not "devise" per se, whereas humans design and devise. AI technologies have altered and improved our ability to design and devise, transforming our machinery for science, drug discovery, engineering, manufacturing, etc. The questions arising around AGI and Superintelligence (SI) can lead to more specific and achievable grand challenges.

- The human-machine symbiosis is a complementary theme to those of Artificial Intelligence. Integration of intelligent machines into human workflows and business processes is a vast opportunity landscape akin to the industrial revolution and introduction of electricity. Interconnection of human cognitive activity and machine processing, via a variety of interfaces (including neural), has great potential across many domains. European Union's Human Brain Project[19] was an ambitious grand challenge in this direction to discover a new model of brain research driven by the integration of data and knowledge. Challenges in this category strive to capture the symbiotic relationship between neural and cognitive models of knowledge that achieve intelligence.

---

[17] Aho, Alfred V., and Jeffrey D. Ullman, *Foundations of Computer Science* (New York: Computer Science Press, 1992), Chapter 1, p. 1.
[18] Simon, Herbert A., *The Sciences of the Artificial*, MIT Press, 1969.
[19] Human Brain Project, "Home," https://www.humanbrainproject.eu/en/.

- Trust, transparency, and security: as computers are our mediators and active participants in society, understanding the limits of trust and security have proven to be an evergreen topic. Challenges in this category have a common core of cybersecurity issues that extend to issues created by the synthesis of AI and human agency, such as the cognitive and societal effects created by disinformation.

These "big vectors" are, on their own, an anticlimax. What we yearn for are specific challenges that are audacious and compelling enough to merit serious study, at scale, by a broad swath of the scientific community — not just computer scientists. We clearly have some work to do to identify what these challenges might be.

## Call to Action

Grand challenges can play an important role in bringing the computing research community together to catalyze innovations that continue to transform science and society. As advances in computing have become essential for addressing the urgent challenges across most disciplines, it is essential that we as a community identify the grand challenges in computing to drive research for the next decade and beyond. Furthermore, it is important that we be intentional about what the desired impacts are and how we will ensure these impacts. Finally we must recognize that the computing innovation ecosystem has evolved, and we must identify necessary research structures and processes (e.g., public-private partnership model) that embrace this ecosystem. This paper is a call to action for the community to work together to identify the next grand challenges in computing.

We encourage those interested in helping move the field forward to:

1. Engage with CCC in discussion on our [LinkedIn](#).

2. Propose a [CCC Blue Sky Track](#) for one of the top conferences in your field.

3. Propose a [CCC Visioning Workshop](#).

4. Think of other ways of engaging your own research community on the question of Grand Challenges and if you do, [let us know by email](#); we would love to hear about it!

## Suggested Citation

# ACKNOWLEDGMENTS

## Reviewers

CCC and the authors acknowledge the reviewers whose thoughtful comments improved the report:

- Gabrielle Allen, University of Wyoming
- David Danks, University of California, San Diego
- Sebastian Elbaum, University of Virginia
- Michela Taufer, University of Tennessee, Knoxville

## U.S. National Science Foundation

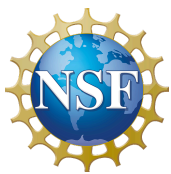

*The CCC AI Futures Task Force was supported by the Computing Community Consortium through the National Science Foundation under Grant No. 2300842. Any opinions, findings, and conclusions or recommendations expressed in this material are those of the authors and do not necessarily reflect the views of the National Science Foundation.*

### About the Computing Community Consortium (CCC)

A programmatic committee of the Computing Research Association (CRA), CCC enables the pursuit of innovative, high-impact computing research that aligns with pressing national and global challenges. Of, by, and for the computing research community, CCC is a responsive, respected, and visionary organization that brings together thought leaders from industry, academia, and government to articulate and advance compelling research visions and communicate them to stakeholders, policymakers, the public, and the broad computing research community.